\documentclass{appolb}
\pdfoutput=1
\usepackage{amsmath}
\usepackage{amssymb}
\usepackage{graphicx,bbm,mathrsfs}
\usepackage{nicefrac}
\usepackage{bbm}
\usepackage{geometry}       
\usepackage{dcolumn}   
\usepackage{bm}        
\usepackage{multirow}
\usepackage{color}

\newcommand{\be}{\begin{equation}}  
\newcommand{\ee}{\end{equation}}  
\newcommand{\bea}{\begin{eqnarray}}  
\newcommand{\eea}{\end{eqnarray}}  

\def\gev{\, {\rm GeV}}

\DeclareRobustCommand{\fbi}{\ensuremath{\mathrm{fb}^{-1}}}

\begin{document}
\title{Prospects for new physics searches at the
LHC in the forward proton mode%
\thanks{Invited talk at EDS Blois 2015: The 16th conference on Elastic and Diffractive Scattering}%
}
\author{Sylvain Fichet
\address{ICTP South American Institute for Fundamental Research, Instituto de Fisica Teorica,\\
Sao Paulo State University, Brazil \\}
}
\maketitle
\begin{abstract}
The installation of forward proton detectors at the LHC will provide the possibility to observe central exclusive processes, opening a novel window on physics beyond the Standard Model. We review recent developments on the  discovery potential from  central exclusive light-by-light scattering. 
The search for this process is expected to provide bounds on a wide range of  particles, and turns out to be complementary from new physics searches in inclusive channels.

\end{abstract}

  \section{Central exclusive processes and forward proton detectors}

The installation of new forward detectors is scheduled at both ATLAS (ATLAS Forward Proton detector \cite{atlas}) and CMS (CT-PPS detector \cite{cms}).
The purpose of these detectors is to measure intact protons arising from diffractive processes at small angle. Among all possible processes, the so-called central exclusive processes  have the structure
\be
pp\rightarrow p \,\oplus X  \,\oplus p\,,
\ee
where the $\oplus$ denote gaps with no hadronic activity between the central system $X$ and the outgoing protons. 
These central exclusive processes can potentially provide a new window on physics beyond the Standard Model (SM) at the LHC. The crucial feature of these processes  is that the whole event kinematics could potentially be known, provided that the invariant mass of the proton system is measured. Such information could then be used to drastically reduce the backgrounds.

\begin{figure}
\begin{center}
\includegraphics[trim=0cm 2cm 0cm 0cm, clip=true,width=0.70\linewidth]{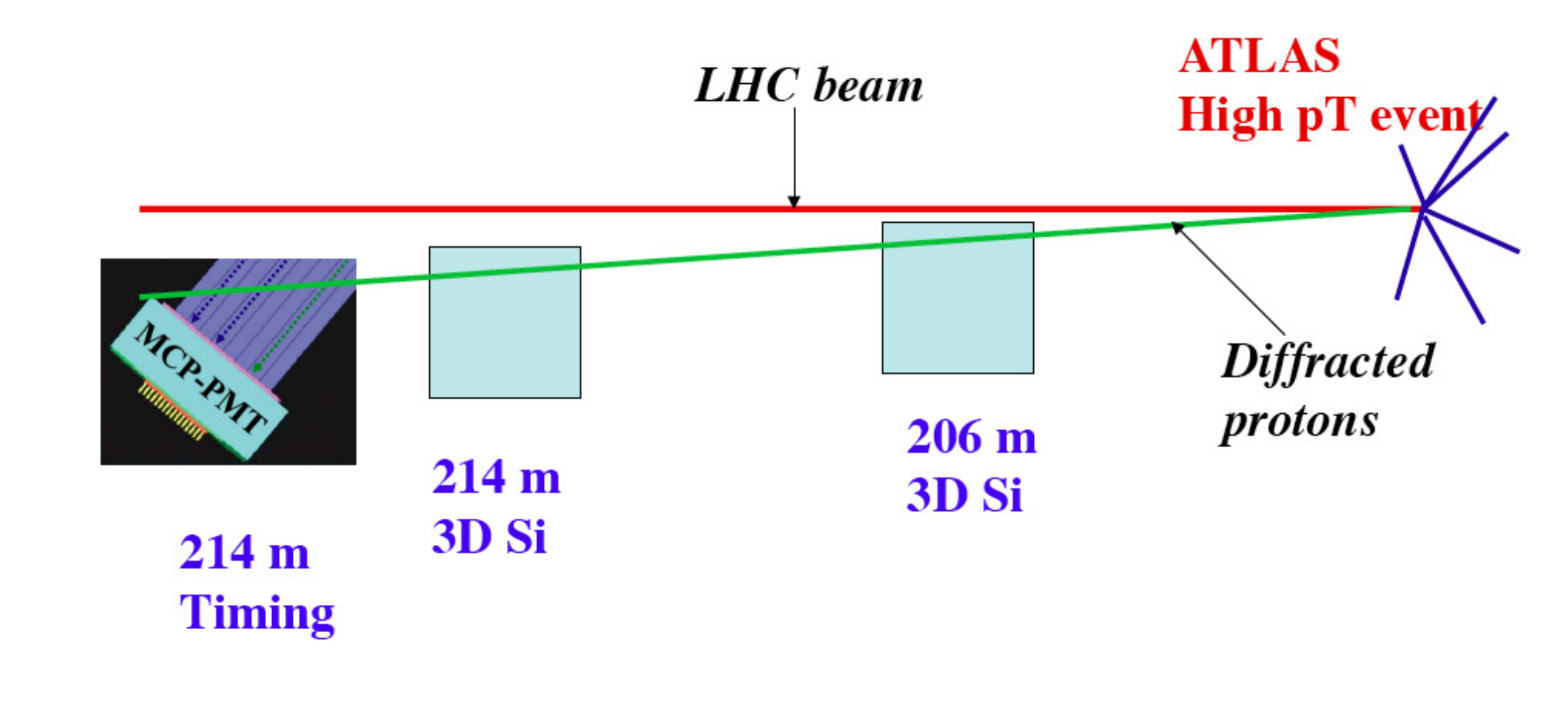}
\end{center}
\caption{Scheme of the AFP detector. Roman pot hosting Si and timing detectors
will be installed on both sides of ATLAS at 206 and 214 m from the ATLAS nominal interaction point. The
CMS-TOTEM collaborations will have similar detectors.}
\label{afp}
\end{figure}

The future forward detectors are precisely designed to perform such characterisation of the outgoing protons.   They will be built at $\sim200$ m on both sides of CMS and ATLAS. The detectors should host  tracking stations, as well as timing detectors (see Fig.~\ref{afp}).
The proton taggers 
are expected to determine the fractional proton momentum loss $\xi$ in the 
range $0.015 < \xi < 0.15$ with a relative resolution of 2\%. 
In addition, 
the time-of-flight of the protons can be measured within 10~ps, which 
translates into $\sim 2$~mm resolution on the determination of the interaction 
point along the beam axis $z$. 

\begin{figure}
\begin{center}
\includegraphics[trim=0cm 0cm 0cm 0cm, clip=true,width=5cm]{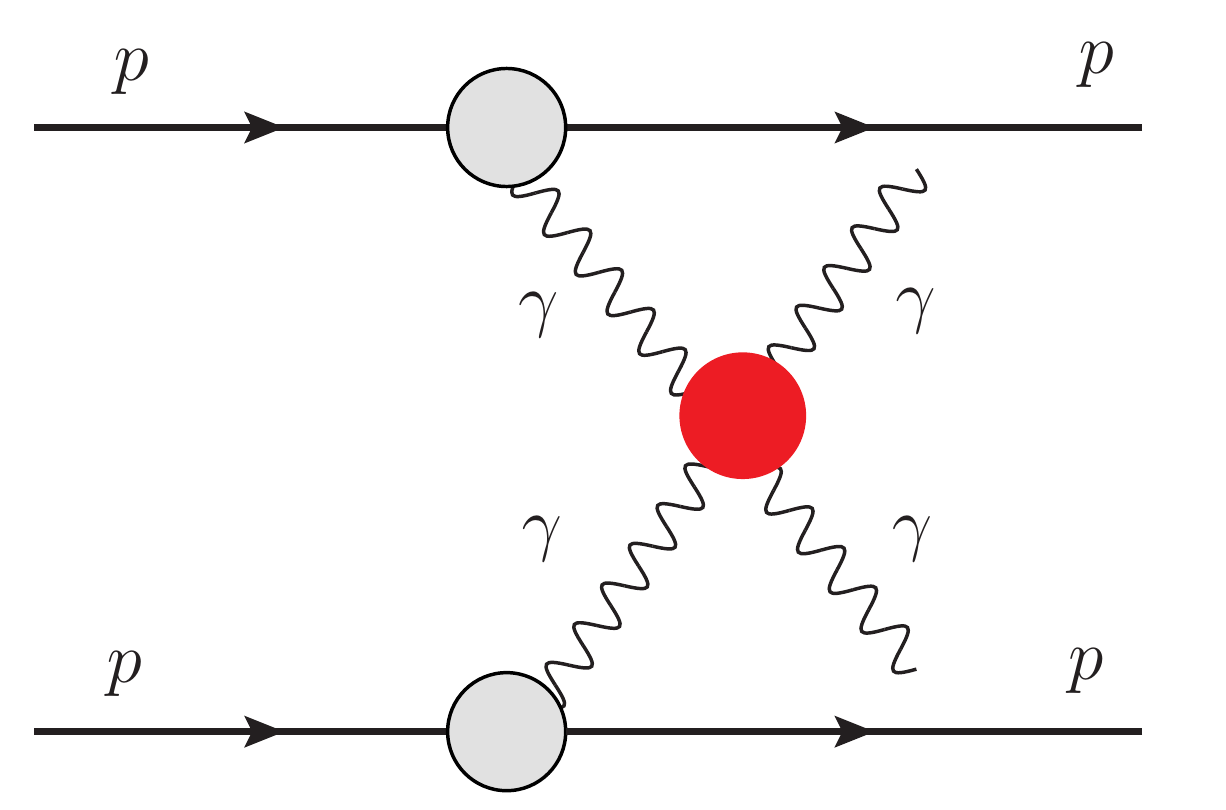}
\end{center}
\caption{Light-by-light scattering with intact protons. }
\label{fig:4gamma}
\end{figure}

 This experimental setup provides a clean environment to look for physics beyond the SM.  
 Central exclusive processes with intermediate photons are the mostly studied ones, because the equivalent photon approximation is well understood. In principle, at the LHC energies, intermediate $W$, $Z$ bosons could 
also happen, however a precise estimation of the fluxes is needed. 
In terms of an effective theory description of the new physics effects,  operators like $|H|^2 V_{\mu\nu}V^{\mu\nu}/\Lambda^2$ induce anomalous  single or double Higgs production (for the MSSM case, see \cite{Heinemeyer:2007tu, Tasevsky:2014cpa}). 
The flavour-changing dipole operators like $\bar{q}\sigma_{\mu\nu} t V^{\mu\nu} /\Lambda^2$ induce single top plus one jet production (see \cite{Fichet:2015oha}). Finally, the four-photon operators of Eq.~\ref{zetas} induce light-by-light scattering. 
This last process is pictured in Fig.~\ref{fig:4gamma}.
 Such self-interactions of neutral gauge bosons are particularly appealing to search for new physics, because the SM irreducible background is small. These interactions constitute smoking gun observables for new physics.
Studies using proton-tagging at the LHC 
for new physics searches can be found in \cite{usww, usw,Sahin:2009gq,Atag:2010bh, Gupta:2011be, Epele:2012jn, Lebiedowicz:2013fta, Fichet:2013ola,Fichet:2013gsa,Sun:2014qoa,
Sun:2014qba,Sun:2014ppa,Sahin:2014dua,Inan:2014mua,Fichet:2014uka,Cho:2015dha}.

\label{se:fwd}

  \section{Measuring light-by-light scattering}

Given the promising possibilities of forward detectors, a  realistic simulation of the search for anomalous $\gamma\gamma \rightarrow  \gamma\gamma$ at the $14$ TeV LHC has been carried out in \cite{Fichet:2014uka}. 
The search for light-by-light scattering at the LHC without proton tagging  has been first tho\-roughly analyzed in \cite{friend:2013yra}.
Let us review the setup, the backgrounds, the event selection, and the sensitivity to the $\zeta_{1,2}$ anomalous couplings expected at the $14$ TeV LHC.

The Forward Physics Monte Carlo  generator (FPMC, \cite{FPMC})  is designed to produce within a same framework the double pomeron exchange (DPE), single diffractive,
exclusive diffractive and photon-induced processes. 
The emission of photons by protons is correctly described by 
the  Budnev flux \cite{Chen:1973mv, Budnev:1974de}, which takes into account the proton electromagnetic structure.
The SM $\gamma\gamma \rightarrow  \gamma\gamma$  process induced by loops of SM fermions and $W$, the exact contributions from new particles with  arbitrary charge and mass, and 
 the anomalous vertices described  by the effective operators Eq.~\eqref{zetas} have been implemented into FPMC.

The backgrounds are divided into three classes. Exclusive 
processes with two intact photons and a pair of photon candidates include the 
SM light-by-light scattering, $\gamma\gamma \rightarrow e^+e^-$ and the central-exclusive production of two 
photons via two-gluon exchange, simulated using ExHume~\cite{ExHuME}.
Processes involving DPE can result in protons 
accompanied by two jets, two photons and a Higgs boson that decay into two 
photons. Finally, one can have gluon or quark-initiated production of two 
photons, two jets or two electrons (Drell-Yan) with intact protons arising 
from  pile-up interactions.

The knowledge of the full event kinematics is a powerful constraint
to reject the  background from pile-up. The crucial cuts consist in matching the missing momentum (rapidity difference) of the di-proton system  with the invariant mass (rapidity difference) of the di-photon system, which is measured in the central detector.  Extra cuts rely on the event topology, using the fact that the photons are emitted back-to-back with similar $p_T$. Further background reduction could even be possible by  measuring the protons
time-of-flight, which provides a complete reconstruction of the primary vertex with a typical precision of 1mm.

 \section{Discovery potential for heavy new physics}

In a scenario where new particles are too heavy to be produced on-shell at the LHC, one  expects that the first manifestations show up in precision measurements of the  SM properties. 
Assuming that the new physics scale $\Lambda$ is higher than the typical LHC energy reach $E_{\rm LHC}$, the correlation functions of the SM fields can be expanded with respect to $E_{\rm LHC}/\Lambda$. At the Lagrangian level, this generates a series of local operators of higher dimension, which describe all the manifestations of new physics observable at low-energy.
 This low-energy effective Lagrangian reads
\be
{\cal L}_{\rm eff}={\cal L}_{\rm SM}+\sum_{i,n}\frac{\alpha_i^{(n)}}{\Lambda^n}\,.
\ee
The effective Lagrangian is somehow the natural companion of precision physics. 
In all generality, the goal of SM precision physics is to get information on the coefficients $\alpha_i^{(n)}$ and the new physics scale $\Lambda$. 
\footnote{
 In order to get meaningful bounds on $\Lambda$, a  statistical subtlety has to be taken into account (see \cite{Fichet:2013jla}), that conceptually boils down to require new physics to be testable. }
For $\Lambda>E_{\rm LHC}$,  four-photon interactions are described by two pure-gauge operators, 
\be
\mathcal{L}_{4\gamma}= 
\zeta_1 F_{\mu\nu}F^{\mu\nu}F_{\rho\sigma}F^{\rho\sigma}
+\zeta_2 F_{\mu\nu}F^{\nu\rho}F_{\rho\lambda}F^{\lambda\mu}
\label{zetas} \,.
\ee
The effect of any   object beyond the SM can be parametrized in terms of the $\zeta_1$, $\zeta_2$ parameters, as well as any experimental search results.

The estimation of  the LHC sensitivities to effective four-photon couplings $\zeta_i$ provided by measuring  light-by-light scattering with proton tagging is performed in \cite{Fichet:2013gsa,Fichet:2014uka}. 
These sensitivities  are given in Table~\ref{sensitivities}
for different scenarios corresponding to medium luminosity 
(300 fb$^{-1}$) and the high luminosity (3000 fb$^{-1}$ in ATLAS) at the LHC. 
It turns out that the selection efficiency is sufficiently good so that the background amplitudes are negligible with respect to the anomalous $\gamma\gamma\rightarrow\gamma\gamma$ signal. A handful of events is therefore enough to reach a high significance. 
The $5 \sigma$ discovery potential as well as the
95\% CL limits with a pile-up of 50 are given.


\begin{table}[t!]

\begin{center}
\begin{tabular}{|c||c|c||c|}
\hline
Luminosity &  300~\fbi & 300~\fbi & 3000~\fbi \\
\hline
 pile-up ($\mu$)  & 50 & 50 & 200 \\
\hline
\hline
coupling (GeV$^{-4}$) &  5 $\sigma$ & 95\% CL & 95\% CL \\
\hline
$\zeta_1$&   $1.5\cdot10^{-14}$ & $9\cdot10^{-15}$  & $7\cdot10^{-15}$\\
\hline
$\zeta_2$&   $3\cdot10^{-14}$& $2\cdot10^{-14}$  & $1.5\cdot10^{-14}$ \\
\hline

\end{tabular}
\end{center}

\caption{5\,$\sigma$ discovery and 95\% CL exclusion limits on $\zeta_1$ and $\zeta_2$ 
couplings in\gev$^{-4}$ (see Eq.~\ref{zetas}). All sensitivities are given for 300 fb$^{-1}$
and $\mu=50$  pile-up events (medium luminosity LHC) except for the numbers of the last column which are given for 3000
fb$^{-1}$ and $\mu=200$  pile-up events (high luminosity LHC). }
\label{sensitivities}

\end{table}
\begin{figure}[h!]
\center
		\includegraphics[trim=0cm 0cm 0cm 0cm, clip=true,width=8cm]{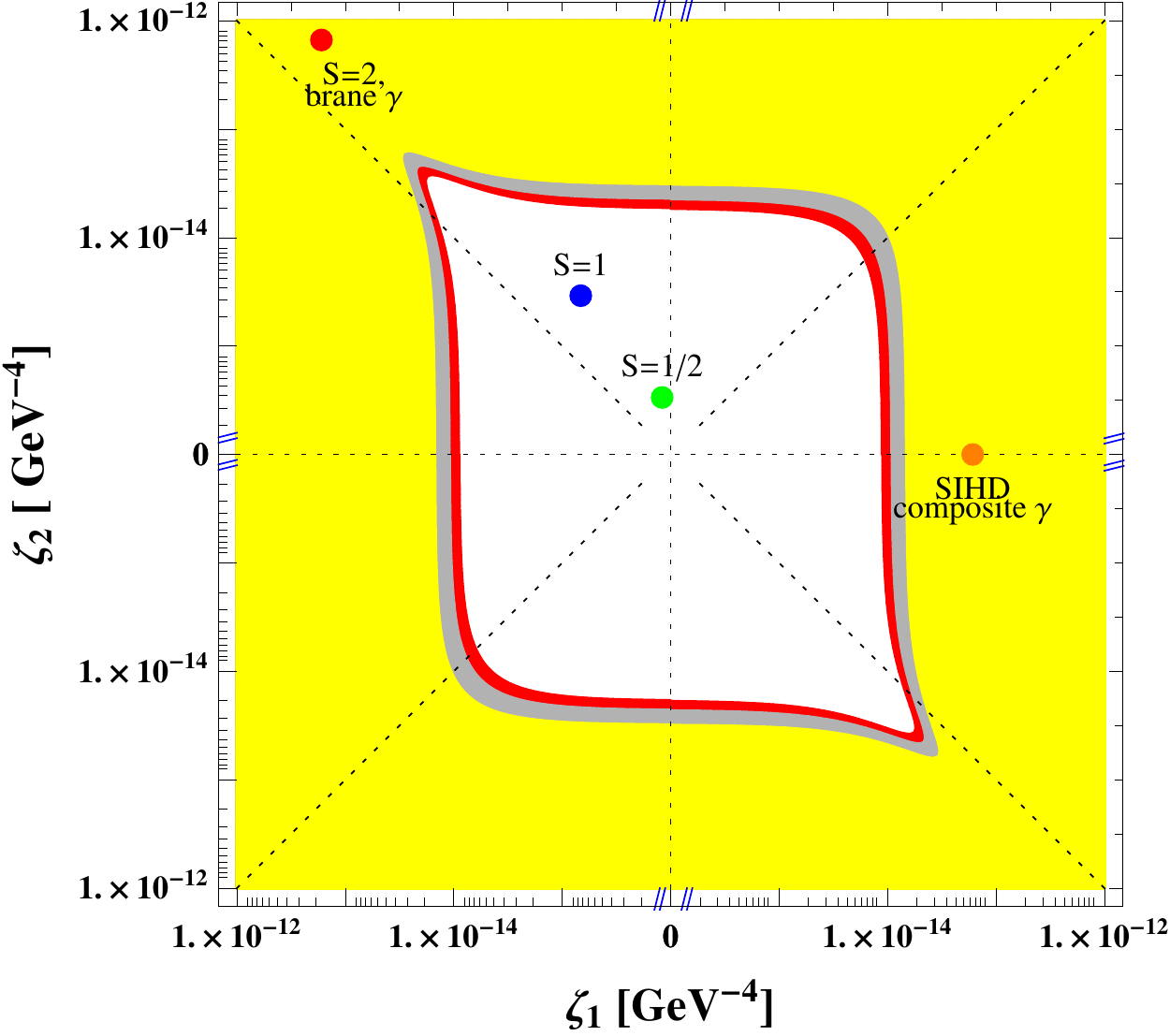}
	\caption{ Experimental sensitivity and models in the EFT parameter space. Axes follow a logarithmic scale spanning $|\zeta_i|\in[10^{-12},10^{-16}]$.
	The yellow, grey, and red  regions can be probed at 5\,$\sigma$, 3\,$\sigma$ and 95\%~CL using proton tagging at the LHC, while 
	the white region remains inaccessible. 	The limits are given for the medium luminosity LHC with all photons (no conversion required) and no form-factor (see Tab. \ref{sensitivities}).	
	Also shown are contributions from electric particles with spin $1/2$ 
	and $1$, charge $Q_{\rm eff}=3$, mass $m=1$ TeV,  
	the contribution from warped KK gravitons with mass $m_{\rm KK}=3$ TeV, 
	$\kappa=2$ and brane-localized photon, and the
	contribution from a strongly-interacting heavy dilaton (SIHD) 
	with mass $m_{\varphi}=3$ TeV coupled to a composite photon.
	 \label{fig:zeta_plot}}
	\end{figure}

 \section{Discovery potential for charged particles}
\label{se:charged}

New electrically charged particles contribute to anomalous gauge couplings  at one-loop. 
Because of gauge invariance, these contributions can be parametrized in terms of the mass and quantum numbers of the new particle \cite{Fichet:2013ola}. In the case of four-photon interactions, only electric charge matters. 
 New  particles with exotic electric charges can for example appear 
in composite Higgs model \cite{Agashe:2004rs} or in warped extra-dimension models with custodial symmetry \cite{Agashe:2003zs}.
 The new particles  have  in general a 
multiplicity $N_{\rm em}$ with respect to electromagnetism. For instance, the multiplicity is
three if the particles are colored. 
It is convenient to take into account this multiplicity by defining
\be
Q_{\rm eff}^4=N_{\rm em} Q^4\,.
\ee

 The SM loops have been computed in Refs.~\cite{Karplus:1950zza,Karplus:1950zz,Costantini:1971cj,Jikia:1993tc} and are collected in Ref. \cite{Fichet:2014uka}. At LHC energies, the $W$ loop dominates over all fermion loops including the top because it grows logarithmically.

\begin{figure}
\begin{center}
\includegraphics[width=0.49\linewidth]{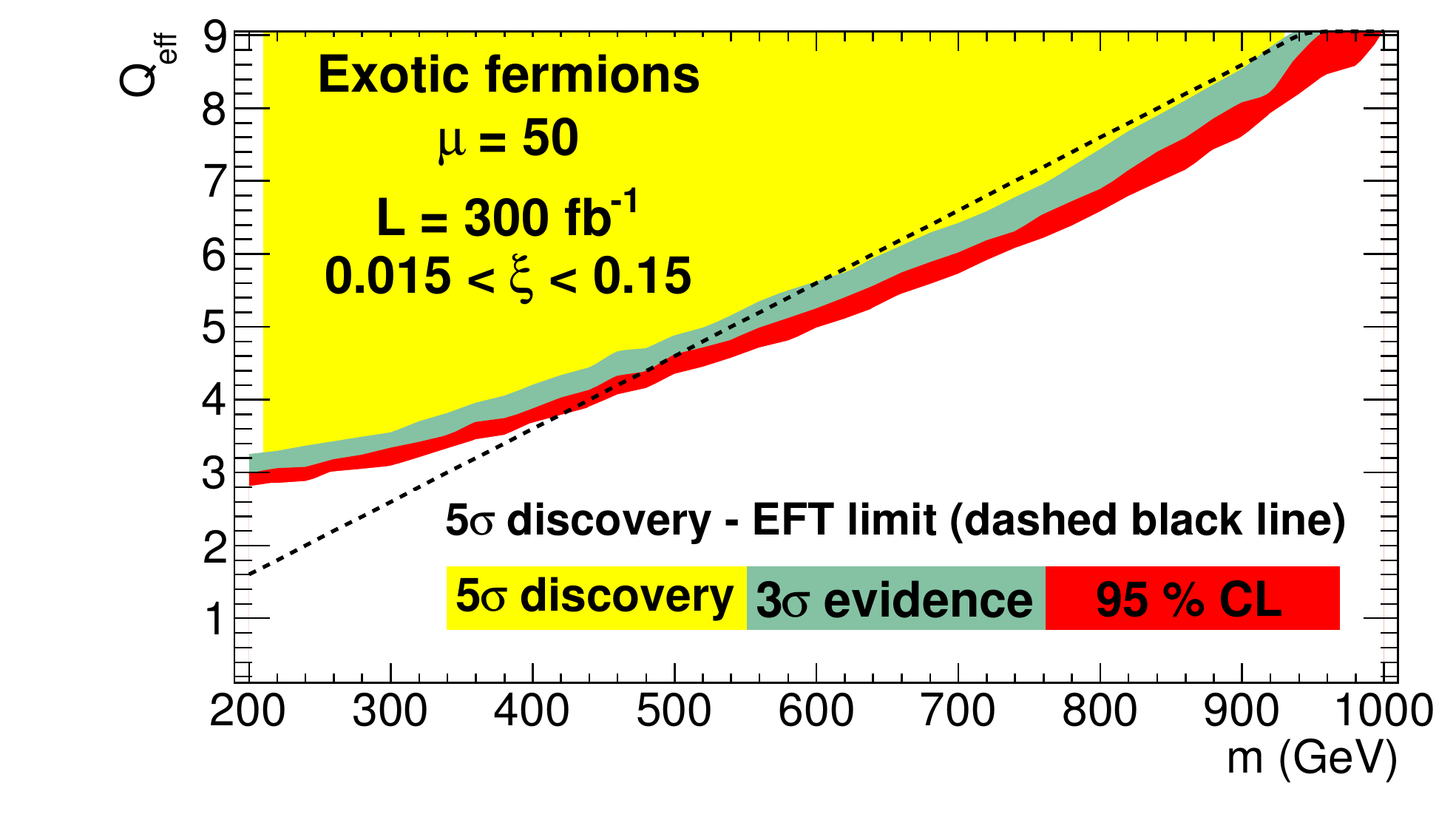}
\includegraphics[width=0.49\linewidth]{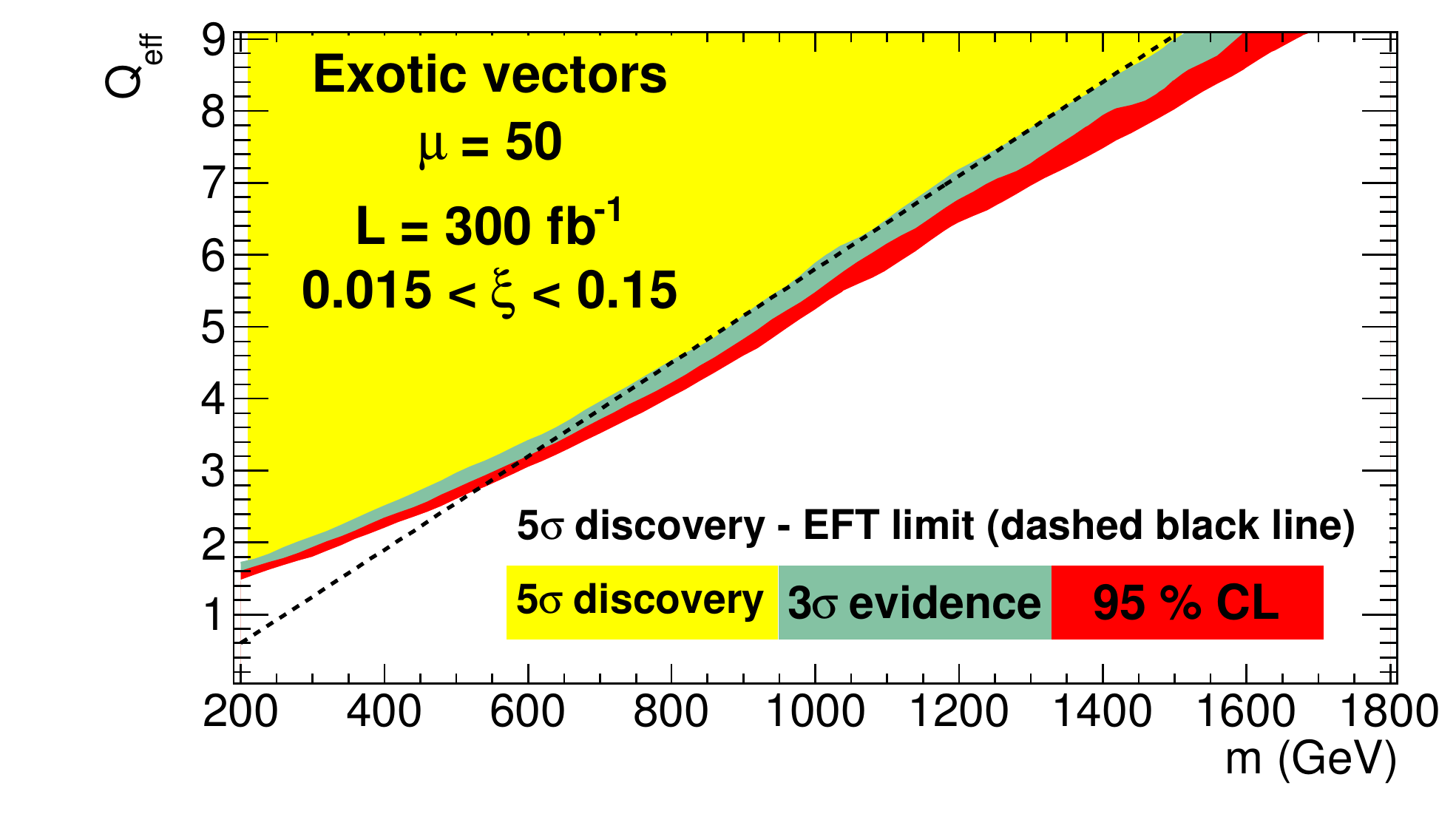}
\end{center}
\caption{Exclusion plane in terms of mass and effective charge of generic fermions and vectors with full integrated luminosity at the medium-luminosity LHC (300~\fbi, $\mu=50$).}
\label{fig:mqplane}
\end{figure}

The  results of the simulation with full amplitudes are given in Tab. \ref{fullamp_values} and
Fig.~\ref{fig:mqplane}  where are displayed the 5$\sigma$ discovery, 3$\sigma$
evidence and 95\% C.L.~limit for fermions and vectors 
for a luminosity of 300 fb$^{-1}$ and a pile-up of 50. 
It is found  that a vector (fermion) with  $Q_{\rm eff}=4$, can be discovered up to mass $m=700$~GeV ($370$~GeV). At high mass, the exclusion bounds follow isolines $Q\propto m$, as dictated by the EFT couplings \cite{Fichet:2013gsa}. Extrapolating the same analysis to a higher luminosity of 3000 \fbi for a pile-up of 200 leads to a slighlty improved sensitivity of  $m = 740$~GeV ($410$~GeV) for vectors (fermions).

One may notice that some searches for vector-like quarks, as motivated from e.g.~Composite Higgs models, already lead to stronger bounds than the ones projected here. For instance, vector-like top partners arising from the $(2,2)$ (corresponding to $Q_{\rm eff}\approx2.2$) of mass $m=500$ GeV would be excluded from present LHC data, while they would be out of reach using light-by-light scattering.
On the other hand, the light-by-light scattering results are completely model-independent. They apply 
just as well to different effective charges, are independent of the amount 
of mixing with the SM quarks, and even apply to vector-like leptons!

\begin{table}

\begin{center}
\begin{tabular}{|c||c|c|c|c|c|}
\hline
Mass (GeV) & 300 & 600 & 900 & 1200 & 1500 \\
\hline
$Q_{\rm eff}$ (vector)  & 2.2 & 3.4 & 4.9 & 7.2 & 8.9 \\
\hline
$Q_{\rm eff}$ (fermion) & 3.6 & 5.7 & 8.6 & - & - \\
\hline
\end{tabular}
\end{center}

\caption{5\,$\sigma$ discovery limits on the effective charge of new generic charged fermions and vectors for various masses scenarios and full integrated luminosity at the medium-luminosity LHC (300~\fbi, $\mu=50$).}
\label{fullamp_values}

\end{table}

 \section{Discovery potential for neutral  particles}

Non-renormalizable interactions of neutral particles are 
also present in common extensions of the SM. Such theories can contain 
scalar, pseudo-scalar and spin-2 resonances, respectively denoted by $\varphi$, 
$\tilde \varphi$ and $h^{\mu\nu}$~\cite{Fichet:2013gsa}, that can be potentially be strongly-coupled  the SM. The full effective theory for such neutral resonances has been recently given in \cite{Fichet:2015yia}. For the coupling to the photon, the most general interactions read 
\be\begin{split}
\mathcal L_{\gamma\gamma}=&f_{0^+}^{-1}\,\varphi\, 
(F_{\mu\nu})^2+f_{0^-}^{-1}\, \tilde\varphi \, F_{\mu\nu}F_{\rho\lambda}\,
\epsilon^{\mu\nu\rho\lambda}/2 \\&+f_{2}^{-1}\, h^{\mu\nu}\, (-F_{\mu\rho} 
F_{\nu}^{\,\,\rho}+\eta_{\mu\nu} (F_{\rho\lambda})^2/4)\,,
\end{split}
\ee
where the $f_S$ have mass dimension one. 

A subtlety, however, is that the width of such resonances  can be very broad. This implies that the momentum dependence of the width has to be kept. From the experimental point of view, this also implies that the standard bump searches become inefficient, and have to be replaced by searches for broad deviations, which are typically more challenging.

A complete study of the discovery potential for broad resonances in central exclusive processes is in preparation (\cite{us_resonances}). 
Here we focus only on the case of the CP-even scalar resonance. 
The complete propagator of such  resonance can be written as
\be
\left<\phi(p)\phi(-p)\right>=\frac{i}{p^2-m^2+i(a\, p^4/(4\pi f_\gamma^{2}) + b p^2 + c)}\,,
\ee
where $a\geq 1$. 
Unitarity remains respected at any energy in this effective theory.  
The case $a=1$, $b=c=0$ corresponds to the minimal case where $\phi$ can decay only through   photons, so that the total width goes as $s^2/f_\gamma^2$. 

\begin{figure}
\begin{center}
\includegraphics[width=0.35\linewidth]{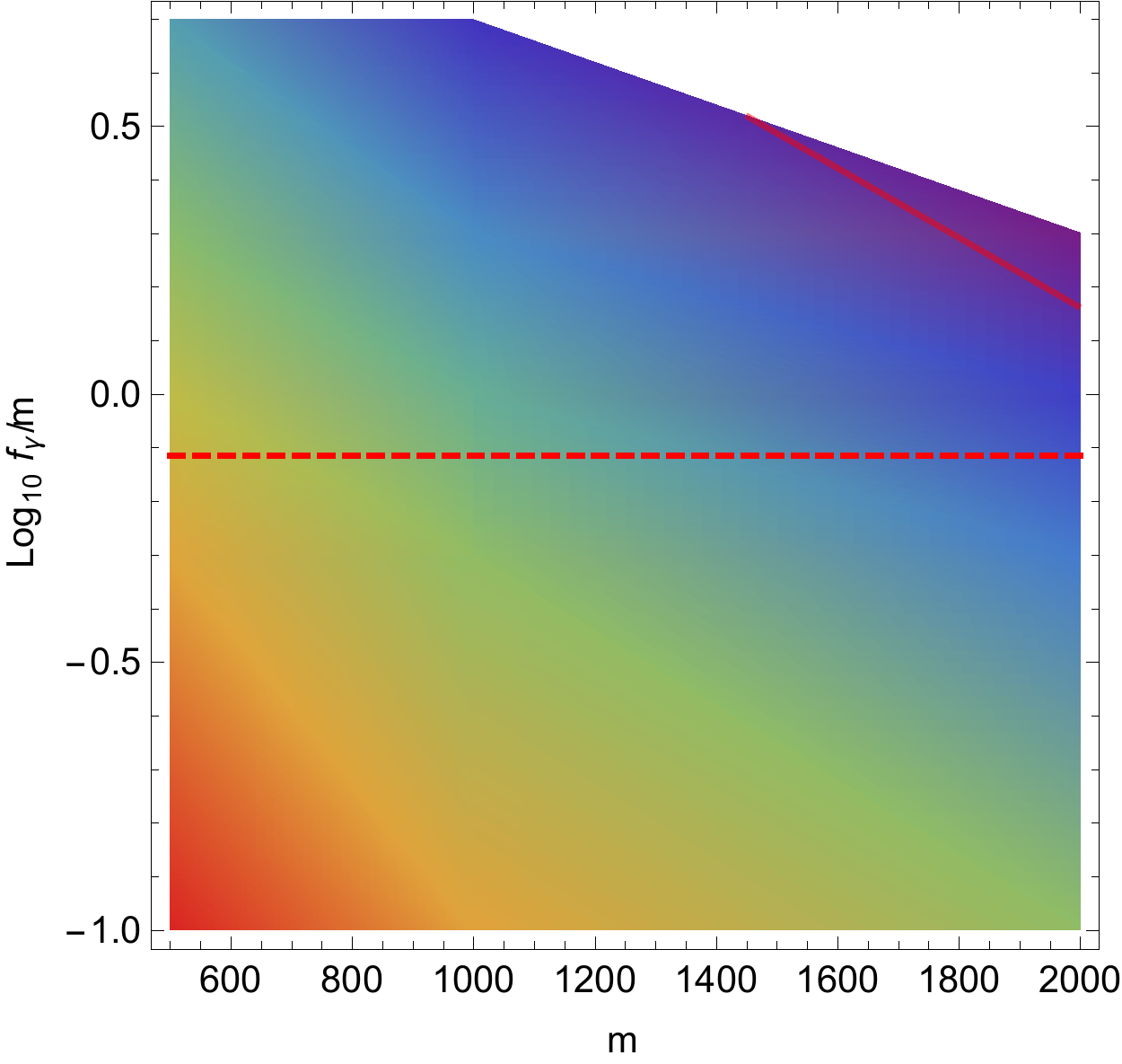}
\includegraphics[width=0.35\linewidth]{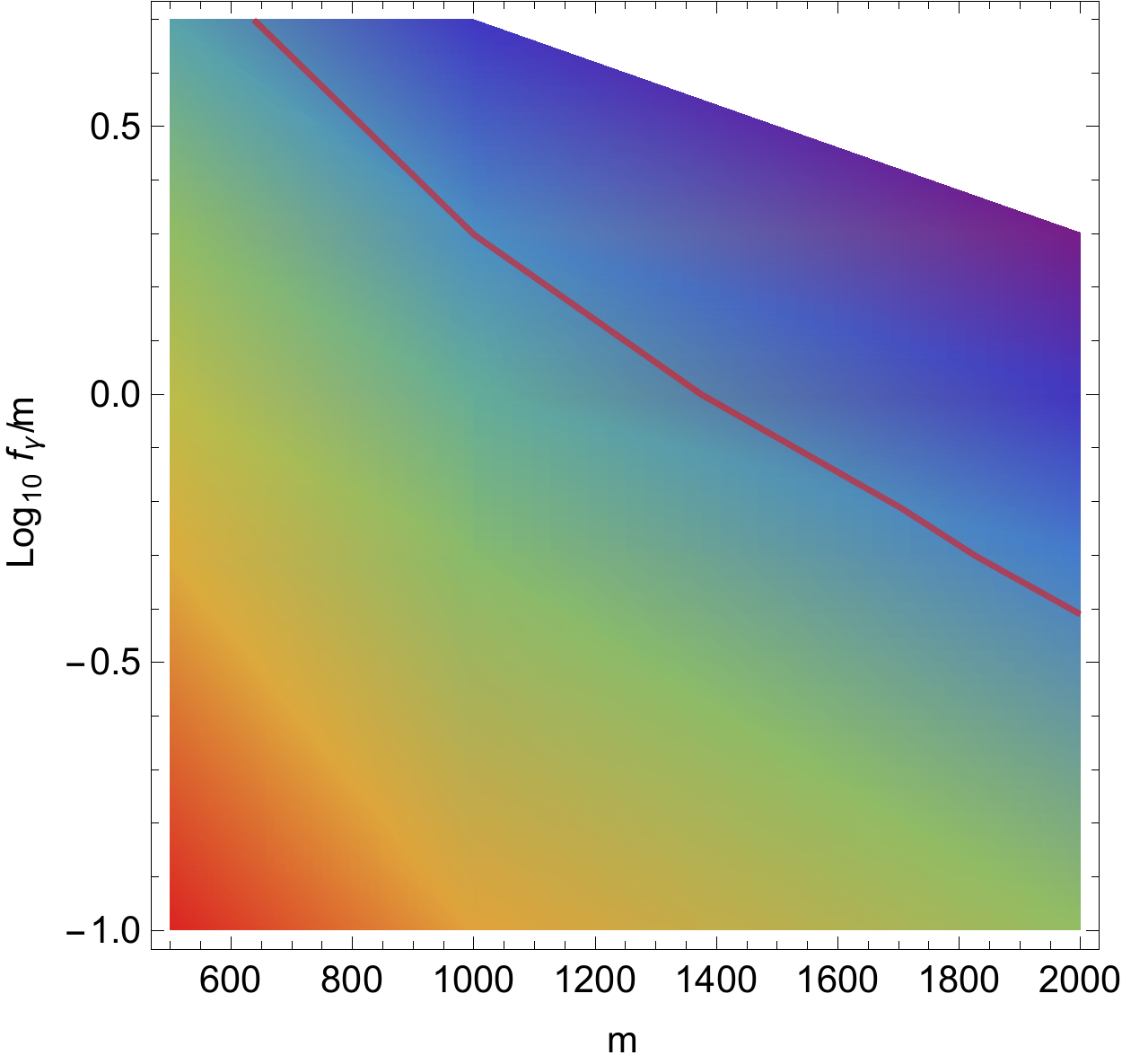}
\end{center}
\caption{Exclusion plane in terms of mass and effective coupling $f_\gamma$ for 
 full integrated luminosity at the medium-luminosity LHC (300\,fb$^{-1}$, $\mu=50$).
Left pannel: $a=1$ . Right pannel: $a=10$. The region below the plain red line can be probed using central exclusive light-by-light scattering. The region above the the dashed line can typically be probed by bump searches in central detectors. 
 }
\label{fig:fmplane}
\end{figure}

In Fig.~\ref{fig:fmplane} we show the discovery potential of the scalar neutral resonance 
through central exclusive light-by-light scattering. As a most optimistic scenario, we let $b=c=0$ and we vary $a$. The region below the thick red line is accessible at $5\sigma$ for 300\,fb$^{-1}$, $\mu=50$. As an indication we also show the limit at which the narrow-width approximation is not valid anymore, such that standard bump searches cannot be applied.

It turns out that central exclusive searches can probe the strong coupling region, while bump searches can probe the weak coupling region. For $a=10$ for example, bump searches cannot be applied at all in the region displayed in Fig.~\ref{fig:fmplane}. We conclude that central exclusive and bump searches are complementary.
However, to obtain a complete picture of the sensitivity from different kind of searches, one should also evaluate the reach from broad resonance searches in central detectors. 
This aspect will be included in a future work \cite{us_resonances}. 

\bibliographystyle{JHEP} 

\bibliography{qgc_JHEP}

\end{document}